# Terahertz-driven ultrafast dynamics of rare-earth nickelates by controlling only the charge degree of freedom

Gulloo Lal Prajapati[1*], Sergey Kovalev[1,2], Igor Ilyakov[1], Atiqa Arshad[1], Gaurav Dubey[3], Ketan S. Navale[4], Dhanvir Singh Rana[3], and Jan-Christoph Deinert[1]

[1]*Institute of Radiation Physics, Helmholtz-Zentrum Dresden-Rossendorf, 01328 Dresden, Germany*

[2]*Department of Physics, TU Dortmund University, 44227 Dortmund, Germany*

[3]*Department of Physics, Indian Institute of Science Education and Research Bhopal, Madhya Pradesh 462066, India*

[4]*Department of Physics, Indian Institute of Technology Indore, Madhya Pradesh 453552, India*

*g.prajapati@hzdr.de

## Abstract

An important strategy for understanding the microscopic physics of strongly correlated systems and enhancing their technological potential is to selectively drive the fundamental degrees of freedom out of equilibrium. Intense terahertz (THz) pulses with photon energies of a few meV, can not only serve this purpose but also unravel their electronic and quantum nature. Here, we present THz-driven ultrafast dynamics of rare-earth nickelates ($RNiO_3$, R = rare-earth atom) – a prototype system to study the Mott insulator-metal transition (IMT). The THz drive of its Mott insulating state induces instantaneous IMT via quantum tunneling of valence electrons across the bandgap while the THz drive of its correlated metallic state leads to overall heating of the conduction electrons. The subsequent relaxations of excited electrons in these two states occur via a two-step process (electron-phonon thermalization and recovery of the charge-ordered insulating state) and a one-step process (electron-phonon scattering), respectively. The relaxation dynamics of the electrons and the absence of acoustic phonon modes, in particular, suggest that the THz photons drive only the charge degree of freedom. The purely electronic, ultrafast and local nature of the THz-induced IMT offers its applications in opto-electronics with enhanced performance and minimal device size and heat dissipation.

## 1. Introduction

With the availability of laser- and accelerator-based intense terahertz (THz) sources, the realization of nonlinear light-matter interaction has extended into the THz regime in recent years, though it is still in its nascent phase.[1-2] The use of intense THz pulses as a pump – driving a system out of equilibrium – has shown diverse applicability, e.g. in the selective drive of soft phonon modes, thereby inducing ferroelectricity,[3] ultrafast switching of superconductivity,[4] high harmonic generation in graphene, topological and Mott insulators,[5-7] manipulating magnetic dynamics,[8-10] etc. Furthermore, THz pulses with photon energies in the meV range are an ideal tool for probing and manipulating low energy interactions present in materials on their characteristic picosecond time scales.[11] Thus, they can be very useful for studying strongly correlated systems where the fundamental charge, spin, orbital and lattice degrees of freedom are strongly coupled, leading to coexistence of multiple phases and phase transitions. Decoupling of fundamental degrees of freedom is essential to gain a microscopic understanding of these phases and phase-transitions. The use of intense THz pulses instead of visible/infrared laser pulses to disentangle the coupled degrees of freedom also significantly reduces the thermal effect, thereby enabling the realization of pure electronic and quantum nature of their interactions.[12-13] By exploiting these features of THz pulses, we study the THz-induced ultrafast dynamics in a strongly correlated system – rare-earth nickelates ($RNiO_3$, R = rare-earth atom) – which are electronic in nature and occur by controlling only the charge degree of freedom.

$RNiO_3$ is a well-known prototype system to study the Mott insulator-metal transition (IMT),[14] which has recently gained increased interest because of the discovery of superconductivity in its layered structure.[15] Some other recently discovered unique properties of nickelates are the coexistence of first- and second-order IMTs,[16] the onset of a metallic phase in its antiferromagnetic (AFM) phase[17] and unusual hysteresis dynamics.[18] The tunability of the IMT in nickelates make them technologically rich and a range of applications have already been demonstrated such as in memristors, modulators, photovoltaics, fuel cells, Mott transistors, neuromorphic computations, etc.[19-24] In its bulk phase-diagram,[25] the first member, i.e., $LaNiO_3$ is a paramagnetic (PM) metal down to low temperatures with a rhombohedral crystal structure. The next two members, i.e., $PrNiO_3$ and $NdNiO_3$ exhibit all the three transitions at the same temperature: IMT, AFM to PM and structural transition from monoclinic to orthorhombic crystal structure. For the remaining members, the magnetic transition is decoupled from the other two transitions. These macroscopic structural, electronic and magnetic transitions simultaneously manifest with charge, spin and orbital orderings and charge/bond disproportionation at the microscopic level.[26-29] THz pulses as an 'observer' have provided important insights about nickelates such as Drude/Drude-Smith like charge carrier transport,[30-31] Fermi/non-Fermi liquid type of metallic state,[32] indication of quantum criticality,[31-32] bad metallic behavior,[31-32] phase-coexistence[33] and manifestation of charge density wave under certain conditions.[34-37] However, the THz field-induced ultrafast dynamics in nickelates and its possible technological applications have not been explored to date.

To study THz-driven dynamics of nickelates, we excited the Mott insulating state as well as the correlated metallic state of nickelates by intense THz pulses and probed the subsequent dynamics with an 800 nm optical pulse – a THz pump optical probe (TPOP) experiment. The THz drive of the Mott insulating state leads to instantaneous IMT via quantum tunneling of charge carriers resulting in doublon-holon pair productions. The subsequent relaxation of the excited electrons occurs in two steps: electron-phonon thermalization (fast relaxation with time constant ~1 ps) followed by recovery of the charge-ordered insulating state (slow relaxation with time constant ~4 ps). In contrast to that, the THz drive of the correlated metallic state leads to heating of conduction electrons i.e., hot carrier generation and relaxation occurs via electron-phonon scattering (with time constant <1 ps). Notably, the THz excitation of the metallic state does not lead to the manifestation of any acoustic phonon mode which is often the case for optical excitations.[38-39] These observations strongly suggest the purely electronic nature of the THz field-driven dynamics of nickelates which is governed solely by the charge degree of freedom. Our study presents a clear picture of how the THz field-induced IMT in nickelates is similar to/different from those induced by other means such as temperature, DC field or photo-irradiation.

## 2. Results and Discussion

**2.1 Experiments:**

We fabricated two nickelate films, each of thickness ~30 nm: $LaNiO_3$ film on both side polished $LaAlO_3$ (100) substrate and $PrNiO_3$ film on single side polished $NdGaO_3$ (001) substrate (see Supporting information, section S1 for thin film growth procedure). The temperature-dependent resistivity data reveals that the $LaNiO_3$ film remains metallic down to low temperature while the $PrNiO_3$ film exhibits a first-order IMT with transition temperature ($T_{IM}$) ~125 K [Fig. 1(a)]. A schematic representation of the TPOP experiment is shown in the inset of Fig. 1(b). The THz pump pulse excites the sample and the subsequent THz field-induced changes in the sample are probed by a time-delayed optical pulse (central wavelength

~800 nm). We performed the TPOP experiment in transmission geometry for the LaNiO$_3$/LaAlO$_3$ (100) film as it was sufficiently transparent to the probe pulse. While for the PrNiO$_3$/ NdGaO$_3$ (001) film, we used reflection geometry because, in this case, the transmission of the probe pulse was significantly low (see Supporting information, section S2 for further details). The change in the transmitted probe intensity upon THz excitation of the sample is quantified as transient transmission $\Delta T = T - T_0$; where $T$ is the transmission with THz pump while $T_0$ is the transmission without the pump. Similarly, in reflection geometry, the transient reflection is defined as $\Delta R = R - R_0$; where $R$ is the reflection with THz pump while $R_0$ is the reflection without the pump. As an example, Fig. 1(b) shows the time-trace of the incident THz field amplitude $E_0$ and corresponding transient optical reflection $\Delta R$ for the Mott insulating state of the PrNiO$_3$ film. We define time zero to be the position where $|\Delta R|$ is maximum ($|\Delta R|_{max}$). We observe that as soon as the THz pulse excites the sample, $\Delta R$ starts dropping. This drop keeps on increasing as long as the THz pulse is present on the sample. Subsequently, an exponential rise in $\Delta R$ over a few picoseconds occurs, followed by a quasi-stationary value of $\Delta R$ at longer time-delays. The dip at ~16 ps is due to the re-excitation of the sample by the back-reflected THz pulse from the underlying substrate.

## 2.2 THz drive of Mott insulating state

**2.2.1 Metallization of Mott insulating state via quantum tunneling:** Fig. 2(a) presents the THz-induced dynamics of the Mott insulating state of the PrNiO$_3$ film. The 2D-plot shows $\Delta R$ as a function of pump-probe delay ($t$) and THz field amplitude ($E_0$), measured at ~26 K. At low $E_0$, $\Delta R$ is negligibly small. However, above a threshold value of $E_0$, a noticeable drop in $\Delta R$ is observed. This drop occurs because, upon THz excitation, the strongly localized charge carriers in Mott insulating state become relatively free. Subsequently, they absorb a fraction of the probe intensity resulting in the drop in $\Delta R$. We define this process as metallization of the Mott insulating state. The magnitude of the drop in $\Delta R$ at $t = 0$ serves as a measure of the metallization. The metallization of Mott insulating state is instantaneous on the THz timescale and monotonously increases with increasing $E_0$ up to the maximum applied THz field, $E_{max} \sim 726 \; kV/cm$. This can be seen more clearly in Fig. 2(b), which shows $|\Delta R|_{max}$ as a function of $E_0$. While the $|\Delta R|_{max}$ is negligibly small at low $E_0$, it increases rapidly and non-linearly above the threshold $E_0$. Thus, the THz field-induced metallization is highly non-linear in $E_0$.

There could be multiple possible mechanisms for the observed THz field-induced metallization of the Mott insulating state: multiphoton absorption, resonant excitation of phonon modes, impact ionization and quantum tunneling. The IMT due to multiphoton absorption by valence electrons cannot be favorable because the charge-transfer gap (~1 eV) is more than two-orders of magnitude higher than the THz photon energy (≤ 7 meV).[40] Further, the $|\Delta R|_{max}$ in such case is observed to vary linearly with the fluence above the threshold, while in our case, it is highly non-linear (see further note in Supporting Information section S2).[39, 41] Nickelates have various phonon modes such as vibrations of the Nd ion against the oxygen octahedral corresponding to an energy ~23 meV, a breathing phonon mode at ~62 meV, a vibrational mode at ~85 meV, and the overtone intensities at ~130 meV.[42-43] Clearly, the resonant excitation of any of these phonon modes, which could lead to metallization, is not possible with our THz source. Impact ionization where highly accelerated charge carriers under strong electric field free the bound charge carriers by successive collisions, can also lead to metallization. However, impact ionization in Mott insulators is observed to occur at electric fields on the order of a few MV/cm which is significantly higher than the highest field strength applied in our experiments.[44] Another mechanism for THz field-induced metallization is dielectric breakdown due to quantum tunneling of valence electrons across the gap. This process requires a significantly lower driving field compared to impact ionization.[12-13] In fact, impact ionization is essentially a next step process where the tunneled charge carriers themselves generate additional charge carriers by collisions with bound charge carriers.

We argue that the quantum tunneling of charge carriers from the O-*2p* valence band to Ni-*3d* upper Hubbard band is the most likely mechanism for the observed THz field-induced metallization. Quantum tunneling is analogous to Zener tunneling which is observed in band insulators or semiconductors.[45] Here, at sufficiently high electric field, the charge carriers gain enough kinetic energy to tunnel from the valence band to the conduction band. Thus, above a threshold field strength $E_{th}$, the charge carrier density in the conduction band begins to increase rapidly and the insulator/dielectric becomes conducting. Analogously, in Mott insulators, a strong THz electric field creates doublon (doubly occupied site)-holon (empty site) pairs via quantum tunneling, which hop from site to site leading to electrical conduction [see Fig. 2(c)].[46] In the energy-band picture, this process can equivalently be represented by inclining the valence and conduction bands (for nickelates, they are O-*2p* band and Ni-*3d* upper Hubbard band, respectively). As a consequence, the charge carriers become energetically degenerate allowing them to pass from the valence band to the conduction band. The charge carrier density in the conduction band due to quantum tunneling, can be given by

$$n = aE_0 exp\left(-\frac{\pi E_{th}}{E_0}\right) \quad (1)$$

where $n$ is the charge carrier density, $a$ is a constant and $E_{th}$ is the threshold THz electric field.[12] The increase in $n$ with $E_0$ would lead to increasing absorption of optical photons and hence, a drop in $\Delta R$ i.e.,

$$\Delta R_{max} \propto n = aE_0 exp\left(-\frac{\pi E_{th}}{E_0}\right) \quad (2)$$

As shown in Fig. 2(b), this relation indeed fits well with $E_{th} \sim 145$ kV/cm. This threshold value corresponds to ~0.18 mJ/cm$^2$ of THz pump fluence in our experiment which is significantly smaller than the reported threshold optical fluences (~0.6 mJ/cm$^2$) for photo-induced IMT in nickelates.[39, 41] It suggests that the Mott IMT via quantum tunneling can be induced with much smaller pump fluence values. In Fig. 2(d), we plot $\Delta R$ as a function of pump-probe delay at different temperatures while keeping the THz field amplitude at its maximum. $|\Delta R|_{max}$ successively decreases as we move from low temperature towards $T_{IM}$. Below $T_{IM}$, insulating and metallic domains coexist [as is apparent from the hysteresis in Fig. 1(a)]. It is also known that a finite number of metallic domains in nickelates survives even at temperatures well below the hysteretic region.[18] Metallic domains increase the charge carrier density in the conduction band and thus, reduce the density of states accessible in the conduction band for the tunneled charge carriers. Since increasing the temperature increases the metallic fraction, the density of states in the conduction band becomes less and less accessible. Hence, $|\Delta R|_{max}$ decreases with increasing temperature.

**2.2.2 Relaxation dynamics:**

Fig. 3(a) shows the temporal evolution of $\Delta R$ at ~26 K and at different THz electric field amplitudes $E_0$. For positive time-delays ($t > 0$), the recombination of doublon-holon pairs and consequently, recovery of the Mott insulating ground state begins. The recovery dynamics occur on two distinct timescales: a fast recovery lasting up to ~3 ps followed by a slow recovery that extends past the measured time windows. As shown in Fig. 1(b), the sample does not recover completely to its initial unexcited state even after 40 ps. Rather, it remains in metastable metallic state. The fraction of the Mott insulating state converted into the metastable metallic state increases with THz field amplitude. However, the maximum applied THz field is not sufficient to completely convert the insulating state into metastable metallic state. Otherwise, a peak in $|\Delta R|_{max}$ at some intermediate value of $E_0$ and then a subsequent drop in $|\Delta R|_{max}$ upon further increase of $E_0$ should have been observed.

We fit the time evolution of $\Delta R$ for $t > 0$ with a double exponential decay function:

$$\Delta R = A\exp(-t/\tau_f) + B\exp(-t/\tau_s) + C \tag{3}$$

where $A$ and $B$ are spectral weights corresponding to fast and slow recovery processes, respectively and $C$ corresponds to the long-lived component. $\tau_f$ and $\tau_s$ are fast and slow relaxation time constants, respectively. The experimental data fits well with the function for different $E_0$ values. The obtained values of $\tau_f$ and $\tau_s$ from the fitting at different $E_0$ are plotted in Fig. 3(b) while those of $A, B$ and $C$ are plotted in Fig. 3(c). The relaxation time constants $\tau_f$ and $\tau_s$ are ~1 ps and ~4 ps, respectively and remain nearly unchanged with $E_0$. Rather, the magnitudes of their spectral weights $A$ and $B$ and the long-lived component $C$ increase with increasing $E_0$. This becomes further clear when we plot normalized $\Delta R$ traces for different $E_0$ where all $\Delta R$ curves nearly overlap with each other [see inset of Fig. 3(a)].

We assign the fast component of $\Delta R$ to electron-phonon thermalization while the slow component to the recovery of charge-ordered insulating state.[38-39, 41] The tunneled charge carriers upon THz excitation, reach into thermal equilibrium by mutual exchange of kinetic energy and momentum via elastic scattering. This process is nearly instantaneous on the THz timescale. On later timescale of ~1 ps, cooling of charge carriers occurs via electron-phonon interactions, leading to phonon thermalization. This process corresponds to the fast component of $\Delta R$. On a longer timescale, the recovery of the charge-ordered insulating state occurs and corresponds to the slow component of $\Delta R$. The overlapping of normalized $\Delta R$ curves with each other suggests that increasing the THz field amplitude mainly increases the number of tunneled charge carriers rather than increasing the kinetic energy of already tunneled charge carriers. This means that, increasing the THz field amplitude does not increase electron-electron and electron-phonon scatterings noticeably. Hence, the variation in $\tau_f$ and $\tau_s$ with $E_0$ is negligible. This is in clear contrast to the optical excitation where both the time constants decrease with increasing the fluence.[39, 41] Further, the value of $\tau_f$ in the THz field-induced IMT is two to three times larger than what is observed in the photo-induced IMT. These findings suggest that during the THz field-driven tunneling of valence electrons across the gap, the lattice degree of freedom remains only weakly perturbed up to the highest applied THz field strength.

**2.3 THz drive of the correlated metallic state**

To investigate THz-driven dynamics of the correlated metallic state of nickelates, we use LaNiO$_3$ film which remains metallic down to low temperatures [see Fig. 1(a)]. Fig. 4(a) shows $\Delta T$ as a function of pump-probe delay at different THz field amplitudes for the LaNiO$_3$ film at ~12 K. Like in Mott insulating state, we observe a sudden change in $\Delta T$ as soon as THz pulse excites the sample. However, as expected, this change is positive. The peak of $\Delta T$ i.e., $\Delta T_{max}$ increases nonlinearly with increasing THz field amplitude $E_0$. At subsequent times, the $\Delta T$ shows a recovery of the sample towards its initial state. Similar to the THz drive of the Mott insulating state, the sample does not reach its initial unexcited state completely within the observed time window. However, this recovery is faster compared to that in Mott insulating state and shows only single exponential decay behavior. Therefore, we fit the time evolution of $\Delta T$ for $t > 0$ ps with single exponential decay function:

$$\Delta T = A exp(-t/\tau) + C \quad (4)$$

which fits well for all $E_0$ values. The obtained values of the fitting parameters are plotted in Fig. 4(b-c). The relaxation time constant $\tau$ is slightly smaller than the fast relaxation time constant $\tau_f$ of the Mott insulating state. Further, unlike the relaxation time constants for THz-excited Mott insulating state, $\tau$ shows a slight increase with $E_0$. Remarkably, there is no oscillatory feature in $\Delta T$ for $t > 0$ which usually appears upon optical excitation of nickelates due to manifestation of acoustic phonon modes.[38-39]

The response of the correlated metallic state of nickelates to the THz excitation is typical of a hot electron gas. The conduction electrons in the correlated metallic state of nickelates, in thermodynamic equilibrium, can be treated as an ensemble of free electrons with large effective mass.[38-39, 41] The THz excitation of metallic state of nickelates then means adding extra energy or heat to these conduction electrons. Thus, upon THz excitation, the temperature of the conduction electrons rises. Consequently, the change in $\Delta T$ is positive and increases with increasing THz field amplitude. Subsequently, these hot electrons thermalize and equilibrate via electron-phonon scattering.[39, 41] As the relaxation dynamics occurs without any phase-transition, a single exponential decay function is sufficient to fit the $\Delta T$ for $t > 0$. The increase in the relaxation time constant $\tau$ with $E_0$ can be attributed to the increase in the charge carrier scattering rate with $E_0$.

The absence of an oscillatory feature in $\Delta T$ for $t > 0$ means that there is no manifestation of an acoustic phonon mode (or strain wave propagation).[38, 47-48] Such an acoustic phonon mode, during photo/THz excitation, can manifest in two ways. In the first mechanism (non-thermal), the spatial redistribution of electrons upon photo/THz-excitation induces an internal electric field which forces cations to displace from their original position. The finite displacement of cations causes mechanical stress in the system resulting in the manifestation of an acoustic phonon mode. In the second mechanism (thermoelastic process), during relaxation after photo/THz-excitation, the electronic subsystem transfers its energy to the phononic subsystem, which increases the lattice temperature. Due to the anharmonicity of the lattice potential, the enhanced lattice temperature produces thermoelastic stress, which again results in manifestation of an acoustic phonon mode. The lattice displacement changes the optical constants of the film which then manifest as oscillations in $\Delta T$ for $t > 0$. Thus, the absence of oscillations in $\Delta T$ for $t > 0$ suggests that both these processes are either absent or very weak. This again indicates that the lattice degree of freedom remains nearly unaffected during the THz excitation of the nickelate films.

### 2.4 THz field-induced IMT vs IMTs induced by other means

Now we compare the THz field-induced IMT with those induced by other means. Due to the simultaneous occurrence of multiple transitions and orderings, the microscopic mechanism of the IMT in nickelates has not been completely understood, yet. The current understanding is that the IMT occurs via a Ni "site-selective Mott transition" without any explicit charge-ordering on the Ni sites.[49-50] In this framework, the electronic state of metallic phase is represented as $d^8\underline{L}$ ($\underline{L}$ is oxygen ligand hole) while in the insulating phase, the alternate Ni sites have $d^8\underline{L}^2$ and $d^8$ configuration corresponding to short and long Ni-O bonds, respectively. Following this mechanism, the IMT in nickelates can be induced both in thermodynamic equilibrium and in non-equilibrium. The Mott IMT induced in thermodynamic equilibrium (e.g., increasing temperature) usually occurs simultaneously with a structural transition. However, recent studies on Mott systems show that the IMT begins with the formation of only metallic domains without the noticeable structural transition.[51-52] The structural transition begins at a slightly higher temperature which ultimately leads, above the $T_{IM}$, to the orthorhombic metallic phase in case of nickelates.

In case of thermodynamic non-equilibrium (e.g. in photo-/THz-induced IMT), the structural transition lags further behind the Mott IMT. In photo-induced IMT, at low to intermediate optical fluences, a metastable metallic state forms without any structural transition.[53-54] This suggests that the optical pulses can selectively drive the electronic degrees of freedom without any direct interaction with the lattice degree of freedom to some extent. However, higher optical fluences lead to IMT coupled with structural transition as well as the manifestation of acoustic phonon modes due to the substantial modulation in the lattice degree of freedom. In this context, THz photons are much superior as they induce IMT by selectively driving only the charge degree of freedom and without leading to the manifestation of acoustic phonon mode. We believe that the employed THz field strength in the present study does not lead to a structural transition as well. However, to confirm this, one needs to perform THz pump structural probe experiment which is beyond the scope of this work.

Finally, we compare the THz field-induced IMT with the DC field-induced IMT. Applying a DC voltage above a threshold value on a Mott insulator induces an IMT via local Joule heating.[55-57] In this mechanism, the temperature of the insulating domains undergoing the transition reaches $T_{IM}$. Thus, these domains simultaneously also undergo a structural transition. The transition of insulating domains begins near the electrodes where the temperature reaches $T_{IM}$ the earliest.[55-57] This takes a few hundred ns (called incubation time). Subsequently, these metallic domains grow and form a filament-like structure along the line joining the two electrodes. On longer timescales, the filament spreads laterally and a larger volume of the sample is converted into metallic phase. Clearly, the DC field-induced IMT is not decoupled from structural transition, the electronic characteristics of the transition is overshadowed by thermal effects and the transition is much slower than the THz field-induced IMT. Further, unlike the THz field-induced IMT, the DC field-induced IMT is global in nature.

## 2.5 Technological perspective

The THz field-induced IMT in nickelates opens a pathway to several straightforward applications in opto-electronics. In the past, the utilization of the ultrafast IMT in nickelates and other correlated oxides in several opto-electronic applications has been proposed and demonstrated such as in optical storage, switches, modulators, tunable photonics, photovoltaics, metamaterial applications, etc.[19-24, 58-59] In the same way, the THz field-induced IMT of nickelates can also be utilized in these applications. Moreover, the purely electronic nature of the THz field-induced IMT and its faster response with a significantly reduced thermal effect compared to DC field- or photo-induced IMT can significantly enhance the speed and performance of these opto-electronic devices.

## 3. Conclusions

We have presented a detailed time-resolved study on THz-driven dynamics of nickelates. The THz drive of the Mott insulating state induces the IMT instantaneously on THz timescales via quantum tunneling of electrons from the valence band into the conduction band. The subsequent relaxation of the THz-excited charge carriers consists of two parts: fast relaxation ($\tau_f \sim 1$ ps) followed by slow relaxation ($\tau_s \sim 4$ ps) corresponding to electron-phonon thermalization and the recovery of the charge-ordered insulating state, respectively. While, THz drive of correlated metallic state leads to hot charge carrier generation. The subsequent relaxation of hot electrons occurs via electron-phonon scattering (τ < 1 ps) and without the manifestation of any acoustic phonon mode. Importantly, all the observed dynamics both in the Mott insulating as well as the correlated metallic states are electronic in nature and the dynamics are controlled by only the charge degree of freedom. This highlights the capability of THz photons to selectively drive a given degree of freedom even in strongly entangled states, thereby shedding light on its role in many-body interactions. Thus, this capability of THz photons can contribute significantly to the understanding of microscopic physics of complex phenomena, enhancing their technological potential and synthesizing light-matter interaction-based novel phases. The purely electronic, ultrafast and local nature of THz-induced IMT in nickelates, in particular, can be utilized in already demonstrated opto-electronic applications with enhanced performance, minimal device size and heat dissipation.

**Acknowledgements**

J.-C. D. and G. L. P. acknowledge support from BMBF Verbundprojekt 05K2022 - Tera-EXPOSE.

**Conflict of Interest**

The authors declare no conflicts of interest.

**Data Availability Statement**

The data that support the findings of this study are available from the corresponding author upon reasonable request.

**Keywords**

THz-driven dynamics, strongly correlated systems, rare-earth nickelates, Mott insulator-metal transition, quantum tunneling, opto-electronic applications

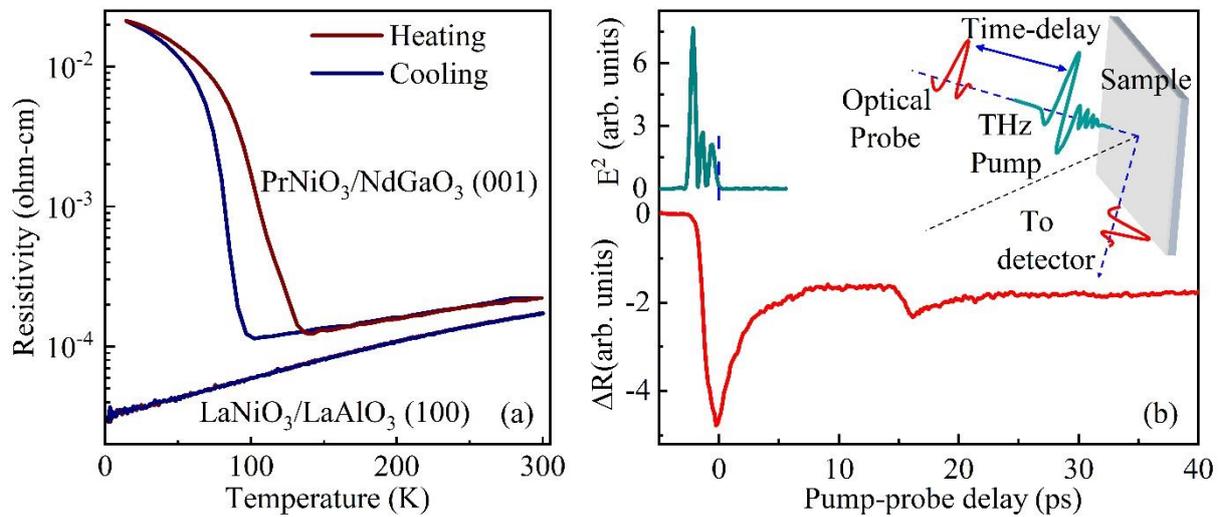

**Fig. 1** (a) Temperature-dependent resistivity of LaNiO$_3$ and PrNiO$_3$ films. (b) Time-trace of THz electric filed amplitude and corresponding transient optical reflection Δ$R$ for PrNiO$_3$ film at ~26 K. Inset shows schematic of THz pump-optical probe experiment in reflection geometry.

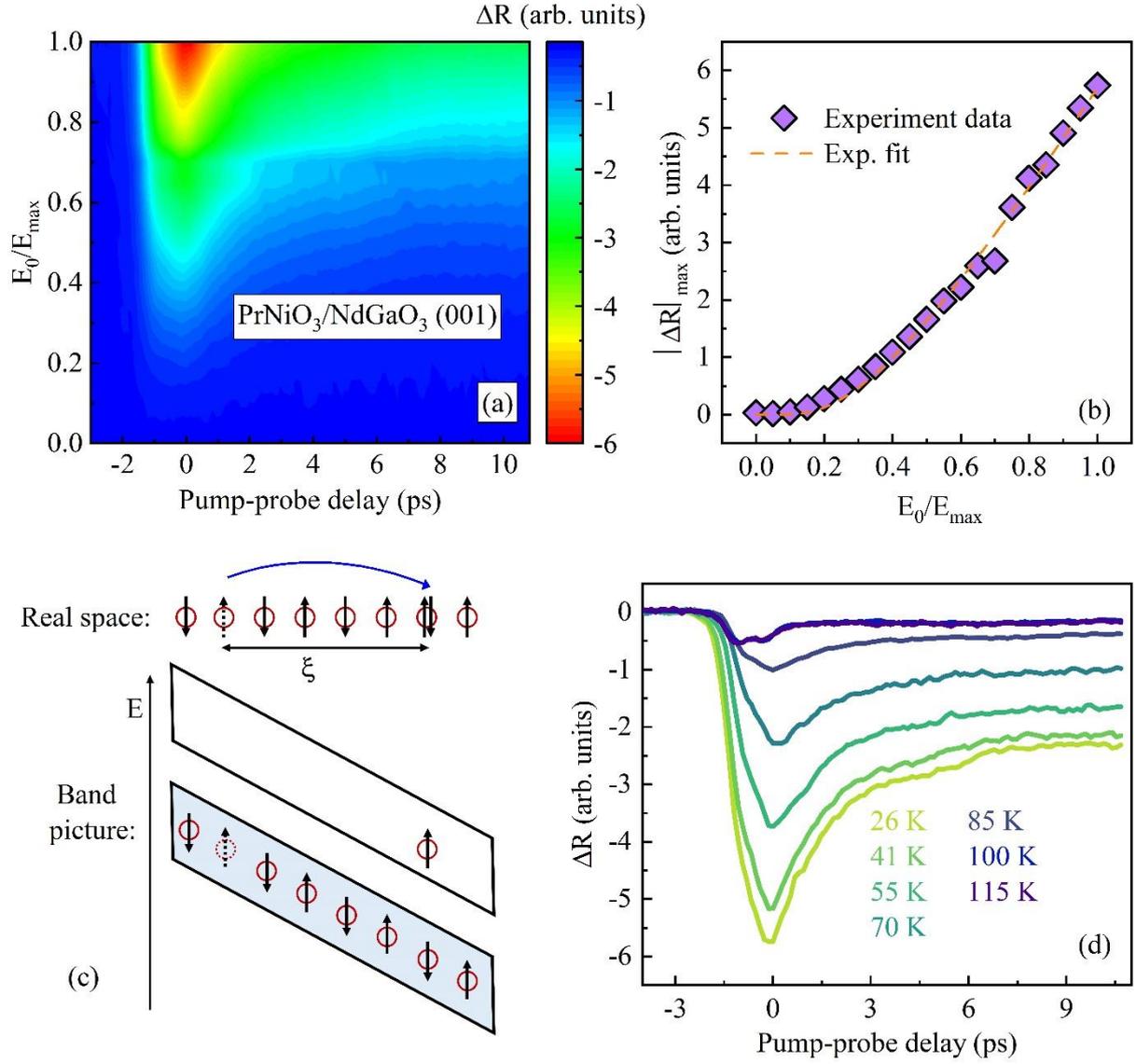

**Fig. 2** (a) Transient optical reflection $\Delta R$ as a function of pump-probe delay and THz electric field amplitude ($E_0$) for PrNiO$_3$ film at ~26 K. (b) $|\Delta R|_{max}$ as a function of THz electric field amplitude. Dashed line is exponential fit (Equation 2) to the experimental data. (c) Quantum tunneling: under strong THz electric field, the charge carriers tunnel through the Coulomb repulsion over a distance ξ (called 'coherence length') producing doublon-holon pairs. This situation in band picture can be presented by making the valence and conduction bands inclined. Inclined bands make the electronic states of electrons degenerate, so the electrons can pass from valence band to conduction band. (d) $\Delta R$ as a function of pump-probe delay at different temperatures while THz pump fluence was kept at maximum.

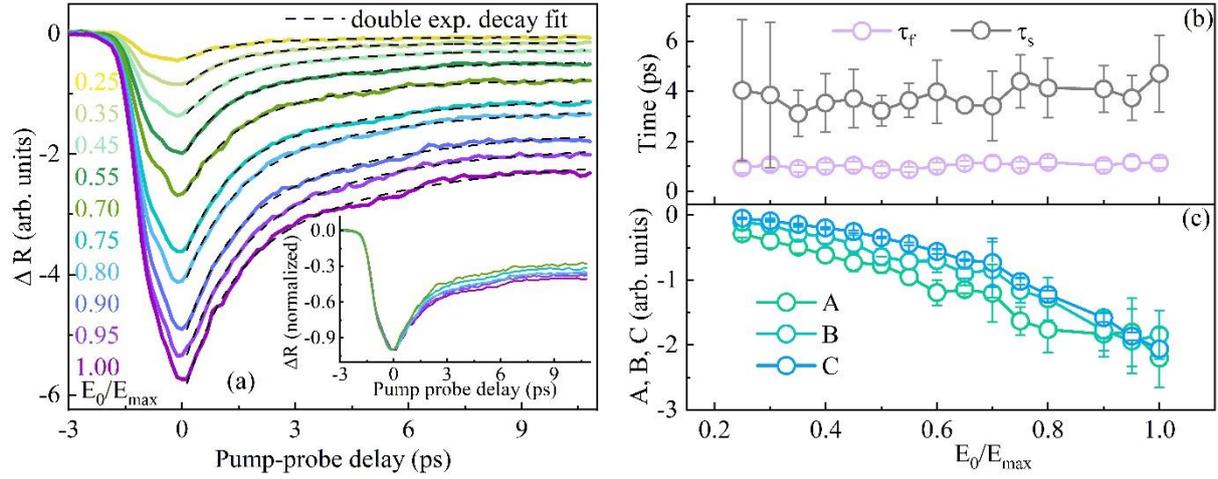

**Fig. 3** (a) Transient optical reflection $\Delta R$ as a function of pump-probe delay at different fixed THz electric field amplitude $E_0$ for PrNiO$_3$ film at ~26 K. Dashed lines are fits to $\Delta R$ for $t > 0$ ps using double exponential decay function (Equation 3). Inset shows $\Delta R$ at different $E_0$ after normalization. (b) Fast ($\tau_f$) and slow relaxation time constants ($\tau_s$) extracted from the fits. (c) Spectral weights A, B and C corresponding to fast, slow and long-lived components of $\Delta R$.

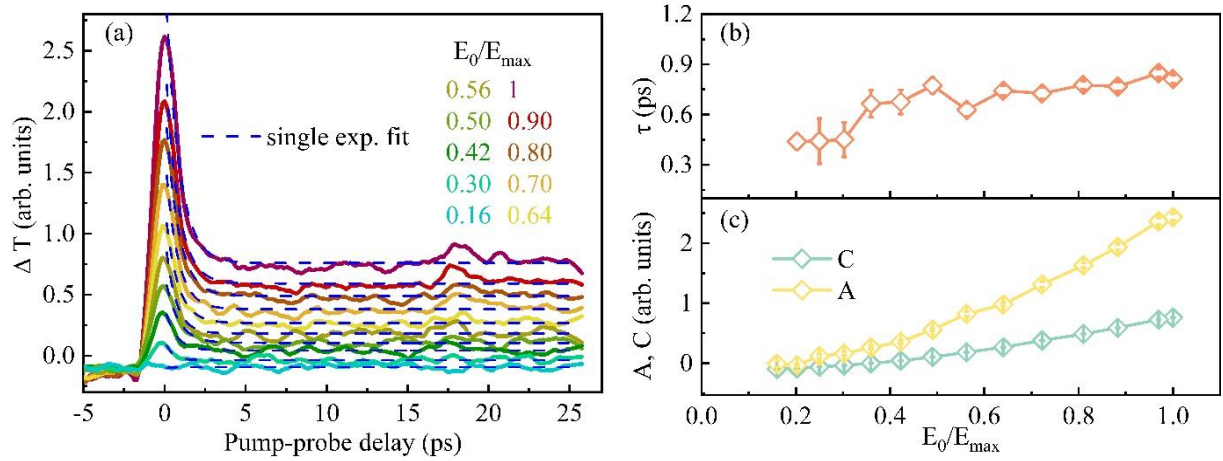

**Fig. 4** (a) Transient optical transmission Δ$T$ as a function of pump-probe delay at different fixed THz electric field amplitude $E_0$ for LaNiO$_3$ film at ~12 K. Dashed lines are fits to Δ$T$ for $t > 0$ ps using single exponential decay function (Equation 4). (b) Relaxation time constant ($\tau$) obtained from the fits. (c) Spectral weights A and C corresponding to decay and long-lived components, respectively.